\documentstyle[11pt,epsfig,amsfonts,amssymb,amstext]{article}

\DeclareFontFamily{U}{rsfs}{\skewchar\font"7F}
\DeclareFontShape{U}{rsfs}{m}{n}{<-6> rsfs5<6-8> rsfs7<8-> rsfs10}{}
\DeclareMathAlphabet{\mathscr}{U}{rsfs}{m}{n}

\def\ben{\begin{equation}}
\def\een{\end{equation}}
\def\be{\begin{equation}}
\def\ee{\end{equation}}
\def\bena{\begin{eqnarray}}
\def\eena{\end{eqnarray}}
\def\bea{\begin{eqnarray}}
\def\eea{\end{eqnarray}}

\def\f(#1/#2){\frac{#1}{#2}}

\def\Z{{\mathbb Z}}
\def\pa{\partial}

\begin{document}

\begin{flushright}
DAMTP-2004-99\\
hep-th/0409307
\end{flushright}

\begin{center}

\vspace{2cm}

{\LARGE {\bf On the stability of naked singularities} }

\vspace{1cm}

Gary W. Gibbons, Sean A. Hartnoll and Akihiro Ishibashi

\vspace{1cm}

{\it DAMTP, Centre for Mathematical Sciences,
 Cambridge University
\\ Wilberforce Road, Cambridge CB3 OWA, UK}

\vspace{0.5cm}

e-mails: g.w.gibbons, s.a.hartnoll, a.ishibashi@damtp.cam.ac.uk

\end{center}

\vspace{1cm}

\begin{abstract}
We study the linearised stability of the nakedly singular
negative mass Schwarzschild solution against gravitational
perturbations. There is a one parameter
family of possible boundary conditions at the singularity. We give
a precise criterion for stability depending on the boundary
condition. We show that one particular boundary condition is
physically preferred and show that the spacetime is stable with
this boundary condition.
\end{abstract}

\pagebreak

\section{Introduction: Phantoms and Runaways}

The positive mass theorems are generally taken as a triumphs of
modern relativity theory. They establish that, under plausible
physical assumptions, asymptotically flat solutions of the
Einstein equations with physically acceptable matter sources
cannot have negative total mass. There is no shortage of solutions
with negative mass, for example the Schwarzschild solution with
negative mass parameter, but they are typical nakedly singular and
hence assumed to be physically unacceptable. They are not expected
to arise from regular initial conditions. However, this begs the
question of why exactly it is that we are reluctant to consider
solutions with negative mass. The question acquires topicality
from various suggestions by cosmologists that the observed
acceleration of the scale factor of the universe may be due to
`phantom matter', that is some matter, typically a scalar field,
with negative kinetic energy \cite{Carroll, Gibbons1}.

The idea of anti-gravity is, of course, a staple of science
fiction writers. According to Mach \cite{mach} it appears to have
been F\"oppl \cite{fop} who first, by analogy with electrostatics,
explored the idea that gravitational masses could be both positive
and negative. The discovery of dark energy may be said to
establish anti-gravity as a serious subject for scientific
discussion and the Randall-Sundrum scenario I \cite{Randall:1999ee},
with its negative tension branes, has only reinforced the trend of
considering matter with exotic energy momentum tensors.

One problem with negative
masses was pointed out long ago in a beautiful analysis of Bondi
\cite{Bondi}. He drew attention to some special features of
negative masses in general relativity.

Firstly, as a consequence of the weak equivalence principle,
a particle of negative mass falling in a gravitational field
should fall at the same rate and in the same direction as a
particle of positive mass. At the level of the geodesic equations
this is because the mass cancels out from the equations of motion.
Thus, for example, a cloud of negative mass particles, let's call
them ghosts or phantoms, should accrete onto ordinary black hole
of positive mass just like ordinary particles.

On the other hand, a body with negative mass should repel
particles with either negative or positive mass. Again, at the
level of geodesic equations this is easily verified for the
negative mass Schwarzschild solution. Combining this fact with our
first observation, we see that the interesting possibility of a
runaway solution exists in which a positive mass particle is
chased in some direction by a negative mass particle, the combined
system going into a state of constant acceleration. The positive
mass particle attracts the negative mass particle to itself, but
at the same time the negative mass particle pushes the positive
mass particle away. See Figure 1.
\begin{figure}[h]
\begin{center}
\epsfig{file=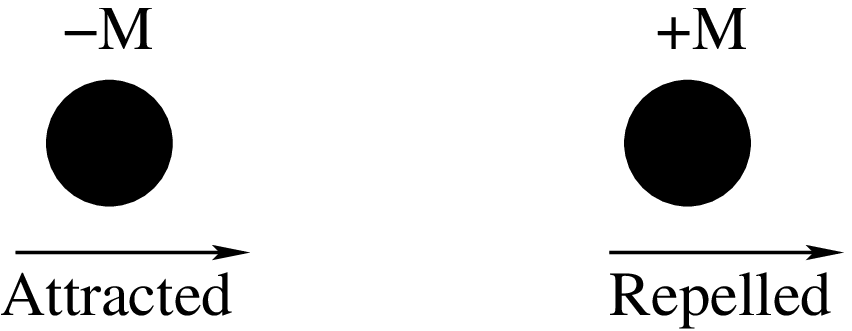,width=6cm}
\end{center}
\noindent {\bf Figure 1.} A negative mass particle chasing a positive
mass particle.
\end{figure}

Bondi showed that the runaway phenomenon actually arises by
exhibiting exact runaway solutions of the Einstein equations. His
paper was written before the development of black hole
theory, but may readily be adapted to include black holes
\cite{Gibbons2}. Bondi considered the axially symmetric static
vacuum metrics first studied by Weyl
\ben
ds ^2 = -e^{2U} dt ^2 + e^{-2U} \Bigl [e^{2k} (d \rho ^2 + dz ^2 )
+
\rho ^2 d \phi ^2  \Bigr ] \,.
\een
The vacuum equations are satisfied if the Newtonian potential
$U=U(\rho,z)$ is harmonic with respect to the standard Laplacian
on three-dimensional Euclidean space ${\Bbb E}^3$ with cylindrical
polar coordinates $(\rho,\phi,z)$. The function $k(\rho,z)$ may
then be obtained by quadratures. A physically acceptable solution
must have $k=0$ on any portion of the axis of symmetry to avoid
conical singularities. This places restrictions on the possible
solutions. The general formalism is reviewed in \cite{Dowker}. The
Schwarzschild solution with positive mass $M$ is obtained by
taking for $U$ the Newtonian potential of a uniform rod of length
$L=2M$. See Figure 2. The portion of the line $\rho=0$ occupied by
the rod is then a regular event horizon. The standard
Schwarzschild coordinates are given by a system of confocal
prolate ellipses in the $\rho-z$ plane, with the rod a degenerate
member of the family.

A rod of mass per unit length $1 \over 2$ and of infinite length,
occupying the positive $z$ axis for example, corresponds to a
Rindler horizon. Superposing a positive mass rod along the
negative $z$ axis gives a solution with a conical defect somewhere
on the axis of symmetry. This solution is the well known C-metric.
Again, see Figure 2. These same conical defects arose in the
solutions considered by Bondi which had as sources a continuous
matter distribution and which were meant to model stars. Bondi's
observation \cite{Bondi}, which also applies to the black hole
case \cite{Gibbons2}, was that by superposing a further source in
between the positive mass Schwarzschild rod and the Rindler rod
the conical defects may be eliminated as long as the new source
has negative mass. In this way a uniformly accelerating runaway
solution may be constructed. The simplest negative mass solution
for $U(\rho,z)$ to take is that of a negative mass point particle
which results in a singular solution called the Curzon solution,
but there are many other possibilities. In particular, one could
use the negative mass Schwarzschild solution. This would amount to
taking the Newtonian potential of a uniform rod of mass per unit
length $-{ 1 \over 2}$. This configuration is shown in Figure 2.
\begin{figure}[h]
\begin{center}
\epsfig{file=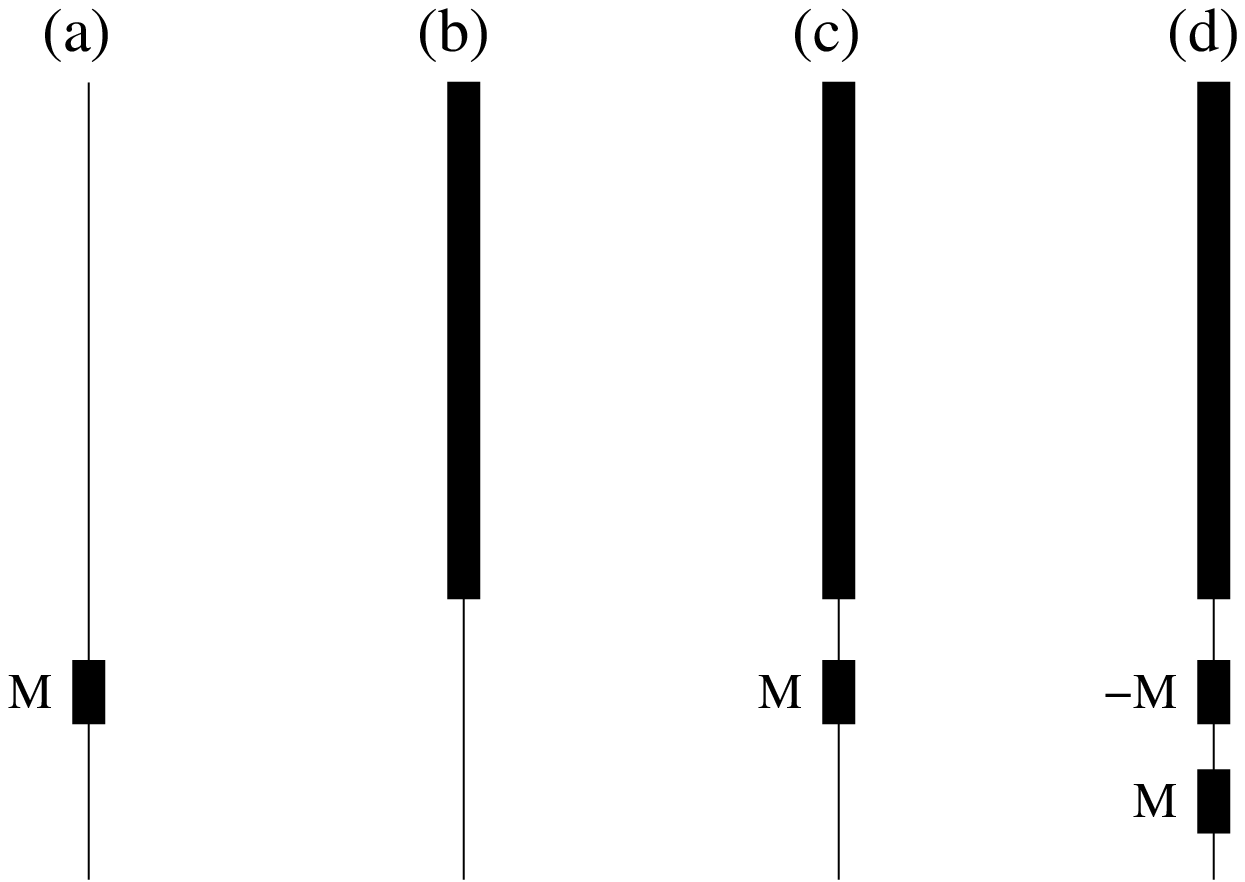,width=8cm}
\end{center}
\noindent {\bf Figure 2.} Rods at $\rho=0$ corresponding to: (a) A
positive mass Schwarzschild black hole, (b) a Rindler horizon, (c) an
accelerating black hole with a conical defect [the C-metric], (d) a
negative mass black hole chasing a positive mass black hole in a
nonsingular spacetime.

\end{figure}

It seems clear from the discussion above that under suitable
circumstances the negative mass Schwarzschild solution should be
unstable. As derived above the instability is dynamical and rather
nonlinear. However, it is an interesting question whether this
instability, or indeed other possibly unrelated instabilities,
appear in a linearised analysis of classical perturbations around the
Schwarzschild phantom. As far as we are aware, a discussion of the
linearised stability of the negative mass Schwarzschild solution
against gravitational perturbations
has not been given before, and the purpose of the present article
is to provide one.

The crucial subtlety in our analysis will concern boundary
conditions at the singularity. The linearised Einstein equations
reduce to a set of Schr\"odinger-like equations with a
time-independent `Hamiltonian'. The choice of boundary
conditions is constrained by the requirement of a self-adjoint
Hamiltonian, and hence a unitary time evolution. For scalar metric
perturbations there is not a unique boundary condition\footnote{The
problem of boundary conditions in negative mass Schwarzschild
spacetime has been studied for test scalar fields in
\cite{Horowitz:1995gi,Ishibashi:1999vw,Giveon:2004yg}.}.
We shall
find a precise range of boundary conditions that guarantee
linearised stability of naked singularity. The existence of
boundary conditions giving linearised stablility of the negative mass
singularity is somewhat counterintuitive. Our results have a very
similar flavour to recent results on the stability of Anti-de
Sitter space, which is also not globally hyperbolic and for which
an ambiguity of boundary conditions exists at infinity
\cite{Ishibashi:2004wx}.

In the Randall-Sundrum scenario I \cite{Randall:1999ee}, the
negative tension brane is stabilised against increasing its area
by a $\Z_2$ quotient of
spacetime that fixes the brane. A similar phenomenon stabilises
orientifolds in string theory and their supergravity realisation in
terms of the Atiyah-Hitchin metric.
The linearised stability of
negative mass black holes that we have found here does not depend
on any such discrete quotient.

The organisation of this paper is as follows. We review the
Schr\"odinger equations describing perturbations of the spacetime.
There is seen to be a one dimensional family of possible boundary
conditions at the singularity. We derive a critical boundary
condition separating stable from unstable spacetimes. There is one
particular boundary condition which results in perturbations with
a more physical behaviour than the others. The spacetime with this
boundary condition is stable. We end with a discussion of our
results.

\section{Stability analysis}

\subsection{Negative mass Schwarzschild spacetime}

We will consider linearised perturbations about the
four dimensional Schwarzschild spacetime. We write the metric as
\bena
ds^2 = - f(r)dt^2
       + \frac{1}{f(r)}dr^2
       + r^2 d\sigma_{(2)}^2 \,,
\label{metric:4DSchwa}
\eena
where $d\sigma_{(2)}^2$ is the metric of a unit sphere and
\bena
 f(r)= 1+\frac{\mu}{r} \,.
\label{def:laps-4dim}
\eena
We are interested in the negative mass case, so $\mu = -2M >0$.
The negative mass Schwarzschild spacetime is well known to have a
naked singularity at $r=0$. This is illustrated in Figure 3, which
shows the Penrose diagram for the spacetime.
\begin{figure}[h]
\begin{center}
\epsfig{file=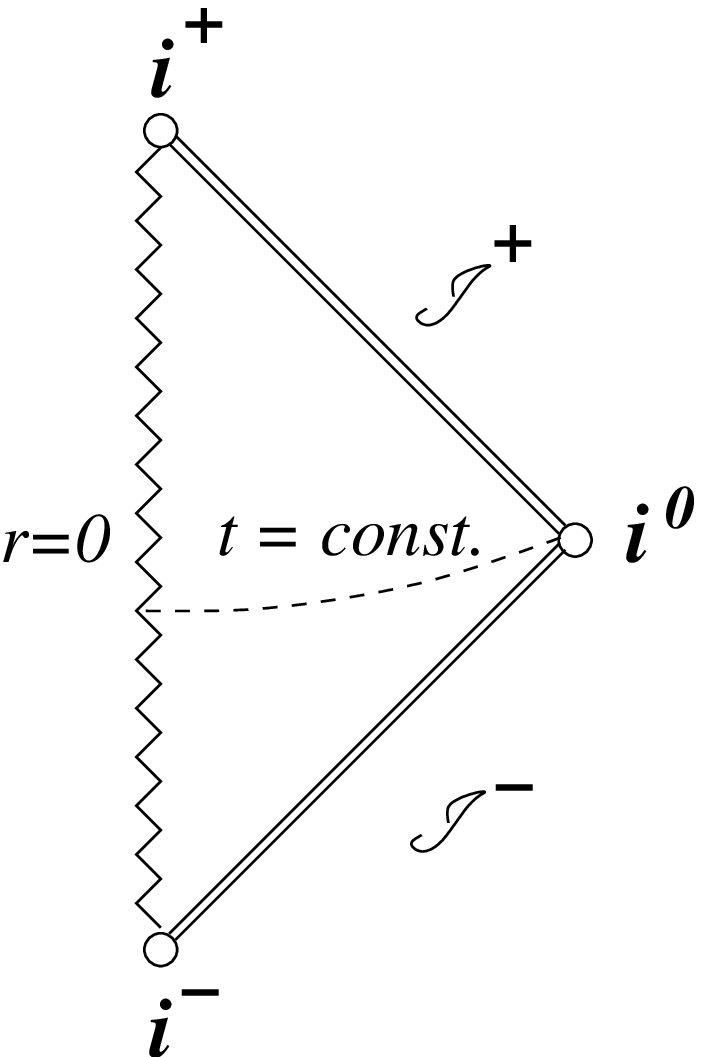,width=2.5cm}
\end{center}
\noindent {\bf Figure 3.} Penrose diagram for the negative mass Schwarzschild:
There is a timelike singularity at $r=0$. The boundary
at $r\to \infty$ consists of two null lines.
Anywhere in this region the Killing field
$\partial /\partial t$ is timelike.
\end{figure}

As usual, it will be convenient to introduce the Regge-Wheeler
coordinate
\ben\label{eq:rw}
r_* \equiv \int \frac{dr}{f(r)}
  = r - \mu \log\left(1+ \frac{r}{\mu}\right)\,.
\een
As $r \rightarrow 0$, $f(r)$ diverges. Therefore the range of the
Regge-Wheeler coordinate is $0<r_*<\infty$. The timelike
singularity is located at $r_*=0$ and infinity corresponds to
$r_*\rightarrow \infty$.

\subsection{Gravitational perturbations}

Gravitational perturbations, $g_{ab} \to g_{ab} + h_{ab}$, of the
four dimensional background metric (\ref{metric:4DSchwa}) may be
grouped into two types: those of axial (vector) and polar (scalar)
type with respect to their parity transformation on the sphere.
Different types of perturbation do not mix at the linearised
level.

In order to solve the linearised Einstein equations, there is a
standard technique for separating the angular coordinates and
constructing gauge invariant scalar variables which we will denote
universally by $\Phi$ \cite{Kodama:2003jz,Ishibashi:2003ap}. The
quantities $\Phi$ are linear combinations of perturbed metric
components and their derivatives: $h,\pa_r h, \pa_{rr}h$. For each
type of perturbation, one obtains an equation of motion for $\Phi$
of the form
\bena
\frac{\partial^2}{\partial t^2}\Phi = \left(
\frac{\partial^2}{\partial r_*^2} - V \right) \Phi \,.
\label{eq:RWZ}
\eena
The potentials appearing in such equations were first derived in a
fixed gauged by Regge-Wheeler \cite{ReggeWheeler} and by Zerilli
\cite{Zerilli}. A unified treatment is given by Chandrasekhar
\cite{chandra,cbook}. The gauge invariant approach has recently
allowed an extension to higher dimensional black holes
\cite{Kodama:2003jz,Ishibashi:2003ap}, including the tensor modes
that only appear in higher dimensions \cite{Gibbons:2002pq}.

The potential function $V$ for the vector/axial/Regge-Wheeler perturbation is
\cite{ReggeWheeler,Kodama:2003jz,chandra}
\ben\label{eq:VofV}
  V_V = \frac{f(r)}{r^2}\left[ \frac{3\mu}{r}+l(l+1)\right] \,,
\een
and for the scalar/polar/Zerilli perturbation the potential is
\cite{Zerilli,Kodama:2003jz,chandra}
\ben\label{eq:VofS}
  V_S = \frac{f(r)}{(mr-3\mu)^2}
  \left[-9 \frac{\mu^3}{r^3}+9m\frac{\mu^2}{r^2}-3m^2\frac{\mu}{r}+ 2m^2+m^3 \right] \,,
\een
with $m=(l-1)(l+2)$ and $l=2,3,4,...$ is the angular momentum.
Note that the $l=0,1$ modes are special.
The former is
spherically symmetric and hence, by Birkhoff's theorem, does not
describe gravitational radiation. Rather it corresponds to a
perturbation changing the mass parameter $M$. The $l=1$ mode is in
fact pure gauge and arises from a translation. We discuss this further
in the conclusion below.

\subsection{Boundary conditions and self-adjoint extensions}

Because of the naked singularity the spacetime fails to be
globally hyperbolic. In general it is far from obvious how to
define the dynamics of any field in such a non-globally hyperbolic
spacetime. However, since the background spacetime we consider is
static we can in fact define sensible dynamics for linear
perturbations. A general prescription for doing so is as follows
\cite{Wald:1980jn,Horowitz:1995gi,
  Ishibashi:1999vw,Ishibashi:2003jd}.

Let $A$ be the spatial derivative part of (\ref{eq:RWZ})
in a non-globally hyperbolic, static spacetime
\bena
A =  - \frac{d^2}{d r_*^2} + V \,, \qquad 0 < r_* < \infty\,.
\label{def:A}
\eena
One may view $A$ as an operator acting on the Hilbert space ${\cal
H}=L^2(r_*,dr_*)$ on a timeslice $\Sigma_t$ orthogonal to the
static Killing field. In order to define unitary dynamics, we need
a self-adjoint extension $A_E$ of $A$. We will see shortly that
choosing a self-adjoint extension corresponds in our case to
choosing boundary conditions at the naked singularity $r_* = 0$.
Given a self-adjoint extension, then the time evolution of $\Phi$
with normalisable initial data $(\Phi_0, \dot{\Phi}_0 )$ on
$\Sigma_0$ is given by
\bena
 \Phi_t= \cos(A_E^{1/2} t) \Phi_0
         + A_E^{-1/2} \sin(A_E^{1/2} t) \dot{\Phi}_0 \,,
\label{def:dynamics}
\eena
where $\dot\Phi_0 = \partial \Phi/\partial t|_{\Sigma_0}$. One
rigorous result that may be proven \cite{Wald:1980jn} is that
whenever the initial conditions are smooth and with compact
support, $(\Phi_0, \dot{\Phi}_0 )\in C^\infty_0(\Sigma_0)
\times C^\infty_0(\Sigma_0)$, then $\Phi_t$ is smooth everywhere
in the spacetime and furthermore within the domain of dependence
of the initial surface $\Sigma_0$, $\Phi_t$ agrees with the
solution to eq.~(\ref{eq:RWZ}) determined from the initial data
$(\Phi_0, \dot{\Phi}_0 )$. Furthermore, it was shown in
\cite{Ishibashi:2003jd} that under certain reasonable conditions
the prescription ~(\ref{def:dynamics}) is the unique way of
defining the dynamics.

Our main interest is in the positivity of $A_E$ or lack thereof.
If $A_E$ is positive then the dynamics is classically stable since
$\cos (A_E^{1/2} t)$ and $A_E^{-1/2} \sin(A_E^{1/2} t)$ in
(\ref{def:dynamics}) become bounded operators. Therefore the time
evolution of $\Phi$ remains bounded for all time and the naked
singularity is stable at the linearised level. On the other hand, if all
possible self-adjoint extensions are not positive, then the
spacetime is unavoidably unstable.

Given a self-adjoint extension $A_E$, one knows that any vector in
the Hilbert space may be expressed in terms of a basis of
eigenvectors of $A_E$. In particular we could prepare smooth
initial data with compact support if we like, even though the
individual eigenvectors of $A_E$ will satisfy neither of these
properties. We therefore turn to a mode analysis of the stability.
A rigorous proof of stability follows from establishing positivity
of $A_E$ \cite{Wald}.

More concretely, consider a mode
\ben
\Phi_t = e^{- i \omega t} \Phi(r_*) \,.
\een
The spatial part then satisfies the Schr\"odinger equation
\ben\label{eq:schrodinger}
A\Phi \equiv - \frac{d^2 \Phi}{d r_*^2} + V \Phi = \omega^2 \Phi \,.
\een
Once we fix the boundary conditions at the singularity, the corresponding
self-adjoint operator $A_E$ will be positive if all normalisable
solutions to (\ref{eq:schrodinger}) have $\omega^2 \geq 0$, and hence
$\omega$ real.

Let us determine the possible boundary conditions. Near the
singularity, the operator for the vector perturbations takes the
form
\ben
A_V \sim - \frac{d^2}{dr^2_*} + \frac{3}{4r_*^2} + \cdots \,, \quad
\text{as} \quad r_* \to 0 \,.
\een
Therefore, the general solution to (\ref{eq:schrodinger})
behaves as
\ben
\Phi \sim a_1 ( r_*^{-1/2} + \cdots ) + b_1 ( r_*^{3/2} + \cdots ) \,, \quad
\text{as} \quad r_* \to 0 \,.
\een
It is immediate to see that both normalisability and
self-adjointness require $a_1=0$. Thus there is a unique
self-adjoint extension $A_E$ of $A$ in this case, which is defined
on the restricted set of functions that satisfy $r_*^{1/2}\Phi
|_{r_*=0}=0$.

The operator for scalar perturbations takes the form
\ben
A_S \sim - \frac{d^2}{dr^2_*} - \frac{1}{4r_*^2} + \cdots \,, \quad
\text{as} \quad r_* \to 0 \,.
\een
In this case the general solution to (\ref{eq:schrodinger}) near the
singularity is
\be\label{eq:0sol}
\Phi \sim a_0 ( r_*^{1/2} \log \frac{r_*}{\mu} + \cdots ) + b_0 ( r_*^{1/2} + \cdots ) \,, \quad
\text{as} \quad r_* \to 0 \,.
\ee
In this case any choice of $(a_0,b_0)\neq (0,0)$ gives
normalisable functions and the corresponding extension of $A$ is
always self-adjoint. From the linearity of the equations involved
it is clear that $(a_0,b_0)$ defines the same self-adjoint
extension as $\lambda (a_0, b_0)$. This equivalence relation
implies that the family of self-adjoint operators is parameterised
by
\be
q \equiv \frac{b_0}{a_0}\quad\in\quad {\mathbb{RP}}^1\,.
\ee
The positivity of $A_E$ will depend on this parameter. A result of
this work will be to show for which values of $q$ the singularity
is stable.

\subsection{A critical boundary condition for stability}

We immediately see that the potential $V_V$ for vector
perturbations (\ref{eq:VofV}) is positive definite. Therefore the
lowest eigenvalue of $A_V$ in (\ref{eq:schrodinger}) is positive
and the spacetime is stable under vector perturbations.

For scalar perturbations the situation is more complicated. Our
strategy is as follows. Firstly we will show that there is a
unique boundary condition, $q = q_C$, such that the minimum
eigenvalue of $A_S$ is precisely zero, $\omega^2 = 0$. This
critical boundary condition separates positive and non-positive
self-adjoint extensions. We will then exhibit boundary conditions
with positive and non-positive spectra on either side of $q_C$.

The most dangerous mode is the $l=2$ mode, as modes with higher
angular momentum are more positive. Figure 4 shows a plot of the $l=2$
potential. For simplicity we will only consider the $l=2$ case in this
subsection.
\begin{figure}[h]
\begin{center}
\epsfig{file=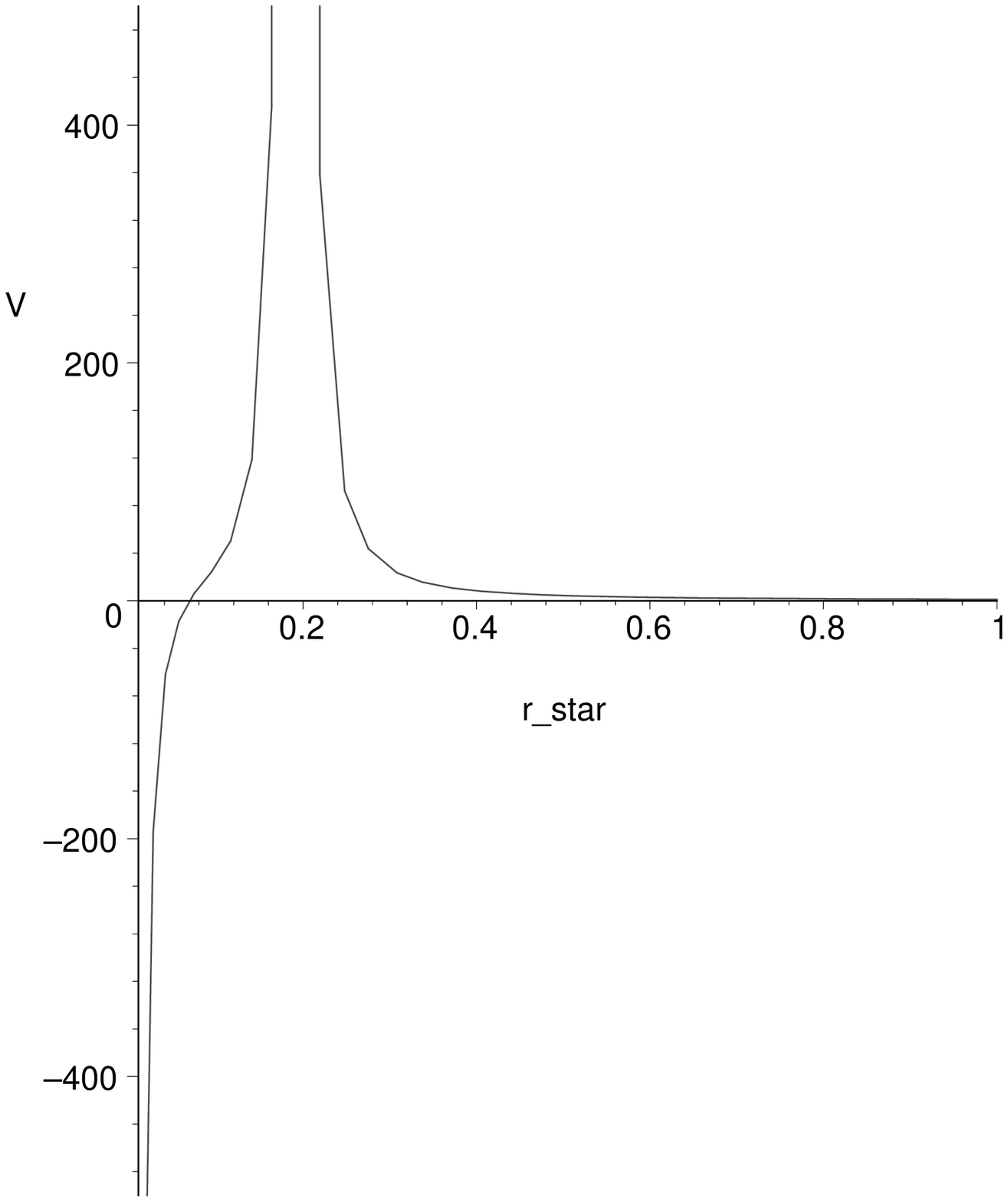,width=6cm}
\end{center}
\noindent {\bf Figure 4.} The potential $V_S(r_*)$ for the $l=2$ scalar
  mode and $\mu=1$ black hole. The
  potential is unbounded below at the singularity and blows up at an
  interior point.
\end{figure}

A curious feature of the scalar potential is that the infinite
centrifugal barrier appears to have been shifted away from the
origin to a finite radius $r_C=3\mu/4$. This divergence is not a
physical effect, but rather an artifact of the variables used to
put the perturbation equations in Schr\"odinger form. We will give
explicit transformations below to curvature variables. The
divergence will have the effect of localising the ground state,
and hence any potential instabilities, near the singularity. This
is reasonable given that the perturbations are required to be
normalisable and that $\mu$ is the only length scale in the system.
One can check that near the singular point the solutions behave as
\be
\Phi \sim C_1 [(r - r_C)^2+\cdots] + C_2 [(r - r_C)^{-1}+ \cdots] \quad \text{as} \quad r \to
r_C \,,
\ee
where $C_1$ and $C_2$ are constants. The solution that diverges as
$r\to r_C$ is not normalisable, so one must impose that $C_2=0$,
that is
\be\label{eq:rc}
\Phi \sim (r - r_C)^2 \quad\text{as}\quad r \to r_C \,.
\ee
This implies that $\Phi =
\Phi' = 0$ at $r=r_C$. Therefore we may truncate the solution at
$r=r_C$ and match it onto any other solution for $r>r_C$ that also
satisfies $\Phi = \Phi' = 0$ at this point. In particular, we may
take $\Phi=0$ for $r>r_C$. This will be the ground state because
$V_S$ is positive for $r>r_C$.

Remarkably, following observations in \cite{Ishibashi:2003ap},
one may explicitly solve the scalar perturbation
Schr\"odinger equation in the case when $\omega^2=0$. One finds
\bena\label{eq:goodsolution}
\Phi & = & C_3 \frac{r(3\mu^3-6r^2\mu+4r^3)}{\mu^3(4r-3\mu)} \nonumber \\
 & + & C_4 \left[\frac{r(13\mu^3-24\mu^2 r +12\mu r^2)}{3\mu^3(4r-3\mu)}
-\frac{r (3\mu^3-6r^2\mu+4r^3)}{\mu^3(4r-3\mu)} \log\frac{r+\mu}{r}\right],
\eena
where $C_3$ and $C_4$ are constants.
We must take the linear combination that does not diverge at $r=r_C=3\mu/4$.
The well-behaved solution has
\be
C_3 = C_4\left[ \log\frac{7}{3}-\frac{4}{9}\right] \,.
\ee
This solution is seen to have no nodes for $r<r_C$. The expression
(\ref{eq:goodsolution}) diverges as $r
\to \infty$. However, as indicated above we can take $\Phi = 0$ beyond
$r=r_C$. Thus we have a normalisable $\omega^2=0$ ground state.
The limit as $r\to 0$ of this solution therefore gives us the critical boundary
condition. Comparing with (\ref{eq:0sol}) we can read off
\be
q_C = 2+\log\frac{98}{9} \,.
\ee
We have at present no conceptual understanding of why there is a
marginal mode and why it occurs at this value of $q$.
The existence of a boundary condition giving $\omega^2=0$
suggests that $q<q_C$ will give ground states with $\omega^2>0$
and hence stable spacetimes and that $q>q_C$ will give unstable
spacetimes. To show this it is sufficient to exhibit one boundary
condition with $q<q_C$ that gives a positive self-adjoint
extension and one negative self-adjoint extension with $q>q_C$.

In fact the situation is a little more complicated because $q$
takes values in ${\mathbb{RP}}^1 \cong S^1$ rather than
${\mathbb{R}}$. There must therefore be at least one other critical
boundary condition at which the minimum eigenvalue changes
discontinuously. Figure 5 illustrates what happens in the simplest
case of one discontinous point. This case occurs for perturbations in
Anti-de Sitter space \cite{Ishibashi:2004wx}.
\begin{figure}[h]
\begin{center}
\epsfig{file=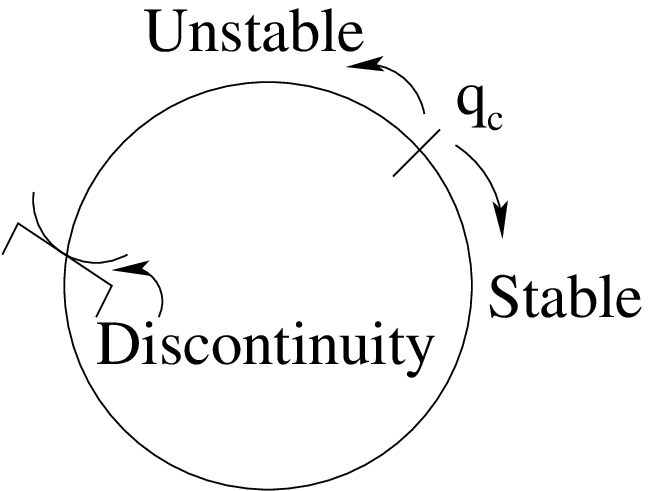,width=3cm}
\end{center}
\noindent {\bf Figure 5.} Stable and unstable boundary conditions
as a function of $q$.
\end{figure}

\subsection{A boundary condition for stability}

In this section we show that the boundary condition $a_0=0$ in
(\ref{eq:0sol}), corresponding to $q=\pm\infty$, gives a positive
self-adjoint extension. The method will be to find a first order
operator of the form
\ben\label{eq:Dtilde}
  \tilde D \equiv \frac{d}{dr_*} + S \,,
\een
with $S$ being some smooth function of $r$.
In terms of this S-deformed operator, we formally obtain
\bena
 (\Phi, A \Phi)_{L^2}
 &=& \left[
           -\Phi^*\tilde D\Phi
     \right]_{r_*=0}^{r_*=\infty}
   + \int dr_*\left(
                   |\tilde D \Phi|^2 + \tilde V |\Phi|^2
              \right) ,
\label{expectvalue}
\eena
where
\ben
   \tilde V \equiv V + f\frac{dS}{dr} - S^2 ,
\een
where $f$ is the metric function (\ref{def:laps-4dim})
and $V$ is the potential in (\ref{def:A}). The expression is formal
because we should check that the boundary term and integral are not both
infinite.

For any smooth function $\Phi$ of compact support the boundary
term vanishes and integration in the second term is finite. If
$\tilde V$ is now shown to be positive for some appropriately
chosen $S$, then the symmetric operator $A_S$ with domain
$C^\infty_0(r_*)$ is positive definite. Indeed we can find a
function $S$ that makes manifest the positivity of the symmetric
operator $A$ for scalar perturbations. We take $S$ to be
\ben
   S = -\frac{f}{r} \,.
\een
Then we have
\bena
  \tilde V_S
    = \frac{f(r)\tilde U(r)}{16 (mr - 3\mu)^2} \,,
\eena
where
\bena
{\tilde U}(r)
 &=&      16 m \left[
                  \frac{\left(
                         m r-3\mu
                   \right)^2}{3r^2}
                  + \frac{2}{3}m^2 +2m
             \right] \,,
\label{tildeQ_S}
\eena
and where as before $m=(l-1)(l+2)$. It is clear that ${\tilde
V}_S$ is positive definite.

Note that $\tilde{V}_S$ is unbounded above at $r=3\mu/m$. For
$l=2$ we have $m=4$ and hence this is the same divergence in the
potential as we found before. As we saw in the previous
subsection, states $\Phi$ must vanish at this point in order to be
square integrable. In this way, we see that $A_S$ is a positive
symmetric operator with domain consisting of smooth functions of
compact support satisfying the regularity condition (\ref{eq:rc})
as $r\to r_C$.

Next we need to extend the domain so that $A_S$ becomes
self-adjoint. This is possible because a positive symmetric
operator always has at least one positive self-adjoint extension,
known as the Friedrichs extension \cite{Wald:1980jn}. We can be
completely explicit in our case: Any self-adjoint extension
corresponding to boundary conditions satisfying
\bena
 \Phi^*\tilde D\Phi \Big|_{r_*=0} \geqslant 0 \,,
\label{BC:singularity}
\eena
will be positive. From (\ref{eq:0sol}) we can see that the
boundary term is zero if $a_0=0$ and diverges otherwise. Thus we
have shown that the boundary condition $a_0=0$ results in a stable
spacetime. This boundary condition corresponds to the Friedrichs
extension, and may be viewed as the simultaneous imposition of
generalised Neumann and Dirichlet boundary conditions in the sense
of \cite{Ishibashi:2004wx}.

\subsection{Some boundary conditions for instability}

We may use the Rayleigh-Ritz trial function method to exhibit
explicitly a range of values for $q$ resulting in unstable
spacetimes. Adapted to our context, one has that for any trial
function $\chi \in {\mathcal{H}}$ we have
\be\label{eq:rr}
\omega^2_{\text{min.}} \leq \frac{(\chi, A_S \chi)}{(\chi,\chi)} \,,
\ee
where $(,)$ denotes the inner product of $L^2(r_*,dr_*)$. We
consider the following set of functions, depending on three
parameters $P,Q,S$
\be
\chi = \left(r-\frac{3}{4\mu}\right)^2 \left[(r+S
  r^2)\log\frac{r}{\mu}+Pr+Qr^2 \right] \quad\text{for}\quad r < \frac{3}{4\mu}\,,
\ee
and vanishing for $r\geq 3/4\mu$. In this ansatz, the boundary
condition at the singularity is specified by $P$.
Substituting the ansatz into (\ref{eq:rr}) we minimise the functional
with respect to $Q$ and $S$ at fixed $P$. Using these minimum values
we see that $(\chi, A_S \chi)$ is negative for $P$ in the range
$[1.850,5.087]$. Translating into values of $a_0$ using the relation
(\ref{eq:rw}) between $r$ and $r_*$ it follows that
\be
q \; \in\; [4.393, 10.867] \quad \Rightarrow \quad \text{Unstable.}
\ee
Note that the critical value we found previously has numerical value
$q_C = 4.389$, so the trial functions give a good approximation to the
exact critical value.

\subsection{Energetics}

In this subsection and the following we will relate the variable
$\Phi$ for the scalar perturbations to curvature perturbations and
to an energy integral. We hope in this manner to clarify the
presence of an unphysical divergence in the Schr\"odinger
potential and to see if there is any sense in which the $a_0=0$
boundary condition is preferred.

One may immediately write down an energy which is always finite
and conserved and furthermore always positive for positive
self-adjoint extensions:
\be
E_0 = (\dot{\Phi},\dot{\Phi}) + (\Phi, A_E \Phi) \,,
\ee
where we use $(,)$ to denote the inner product of $L^2(r_*,dr_*)$
and the dot denotes a time derivative. However, in order to
connect this expression to the usual energies that are considered
in physics one should integrate by parts to obtain a term
quadratic in $\pa_{r_*}\Phi$ rather than $\Phi
\pa_{r_* r_*}\Phi $. It is at this point that the different boundary
conditions produce different behaviours. Let us see what happens
if we consider energies with terms quadratic in $\pa_{r_*}\Phi$.

There are at least two notions of energy that we could consider.
The first is the energy associated with the Schr\"odinger equation
\be\label{eq:E1}
E_1 = \int dr_* \left[ \left| \frac{\pa \Phi}{\pa t} \right|^2 +
\left| \frac{\pa \Phi}{\pa r_*} \right|^2 + V_S \left| \Phi
\right|^2 \right] \,.
\ee
Use ${\mathcal{E}}_1$ to denote the integrand of the previous
expression
\be\label{eq:density}
E_1 = \int {\mathcal{E}}_1 dr_* \,.
\ee
By finding a series expansion for the solutions of the
Schr\"odinger equation (\ref{eq:schrodinger}) one finds that if
$a_0=0$ the energy is finite whilst if $a_0
\neq 0$ then ${\mathcal{E}}_1$ diverges at the singularity
\be
{\mathcal{E}}_1 \sim \frac{|a_0|^2}{r_{*}} \log\frac{r_*}{\mu} +
\cdots \quad \text{as} \quad r_* \to 0\,.
\ee
The corresponding total energy is therefore infinite. This result
might be taken to suggest that $a_0 = 0$ is the more physical
boundary condition. The difference between $E_0$ and $E_1$ may be
understood as arising upon an integration by parts
\be
{\mathcal{E}}_0 = {\mathcal{E}}_1 - \frac{\pa}{\pa r_*} \Phi^*
\tilde{D} \Phi \,,
\ee
where $\tilde{D}$ is given in (\ref{eq:Dtilde}) and
${\mathcal{E}}_0$ is defined analogously to (\ref{eq:density}).
The boundary term is the same that arose in (\ref{expectvalue})
and diverges unless $a_0=0$.

So far the energies we have considered have been somewhat
abstract. However, we can clarify the relation of (\ref{eq:E1}) to
a more commonly considered definition of energy in general
relativity. An expression for the energy of a perturbation about a
background spacetime is given by an integral over a spatial
hypersurface
\be\label{eq:energy}
E_2 = -\frac{1}{8\pi} \int G^{(2)}_{a b} n^a \xi^b d\Sigma_{(3)}
\,,
\ee
where $G^{(2)}_{a b}$ is the quadratic variation of the Einstein
tensor under the perturbation, $n$ is a unit vector orthogonal to
the hypersurface, $n^t=1/f^{1/2}$, and $\xi$ is the timelike
Killing vector $\xi^t=1$. The linear variation will vanish because
the perturbation satisfies the equation of motion.

For the energy to be conserved, there must be no flux of energy
into or out of the singularity. Assuming that the perturbation
dies off sufficiently rapidly at infinity, the change in the energy
with time is seen to be
\be\label{eq:boundary}
\dot{E}_2 = \lim_{r\to 0} \frac{1}{8\pi} \int_{S^2_r} r^2 f(r)^{1/2} G_{rt}^{(2)}
d\sigma_{(2)} \,.
\ee

Using, for example, the formalism developed by Chandrasekhar in
\cite{chandra,cbook}, one can calculate $G_{ab}^{(2)}$ in terms of
the scalar metric perturbations. The linearised equations of
motion for the perturbations are equivalent to the scalar
Schr\"odinger equation. We will not review Chandrasekhar's
formalism here, although we have used it extensively to obtain the
results in this subsection and the next. Considering the $r\to 0$
limit one finds two possible behaviours corresponding to the two
possibilities in (\ref{eq:0sol}). The boundary term
(\ref{eq:boundary}) may be shown to vanish in both cases, so that
\be
\dot{E}_2 = 0 \,.
\ee
Thus (\ref{eq:energy}) provides a conserved energy for the
perturbations. If we evaluate (\ref{eq:energy}) we again find that
the different boundary conditions at the singularity give
qualitatively different results. The $a_0 = 0$ boundary condition
has finite energy, whilst the other cases may be shown to give
\be
{\mathcal{E}}_2 \sim |a_0|^2 \frac{1}{r_*} + \cdots  \quad
\text{as}
\quad r_* \to 0 \,.
\ee
Here ${\mathcal{E}}_2$ is defined analogously to
(\ref{eq:density}). Again, the total energy diverges unless
$a_0=0$. Thus we see that finiteness of the energy is independent
of whether we use $E_1$ or $E_2$. However, the degree of divergence itself
is slightly different. We might therefore ask whether there is any
relation between these two expressions for the energy. Indeed
there is. Using, for instance, Chandrasekhar's formalism
\cite{chandra,cbook} the two expressions may be shown to be
identical up to total derivative terms
\bea
{\mathcal{E}}_2 & \propto & {\mathcal{E}}_1 + \frac{\pa}{\pa r_*}
X \nonumber \\
& \Rightarrow & E_1\, \propto\, E_2 +
\left[\,X\,\right]_0^{\infty}
\,.
\eea
If $a_0 \neq 0$ then
\be
X \sim |a_0|^2 \log^2 \frac{r_*}{\mu} + \cdots \quad \text{as}
\quad r_* \to 0
\,.
\ee
Therefore the two energies are related by a divergent boundary
term.

\subsection{Curvature}

\subsubsection{Curvature scalars}

Given that the background is Ricci flat, the natural curvature
scalar to consider is the Weyl tensor squared. The background has
\be
C_{abcd} C^{abcd} = \frac{12\mu^2}{r^6} \,.
\ee
It is possible to derive an expression for the linearised
perturbation to this curvature scalar in terms of the
Schr\"odinger variable $\Phi$ using, for example
\cite{chandra,cbook}. One finds
\bea\label{eq:curvature}
\lefteqn{\delta \left( C_{abcd} C^{abcd} \right)(r) = \nonumber} \\
& & -12\mu m\left(m+2\right)\left[\frac{\Phi(r)}{r^6} +
\frac{(6r\mu+9\mu^2)f(r)^{1/2}}{(m+2)r^8} \int^r \frac{\Phi(r') dr'}{f(r')^{1/2}(m r'-3\mu)}
\right] e^{-i\omega t} P_l(\cos\theta) \,, \nonumber \\
\eea
where we have restored the angular and time dependence and without
loss of generality we have restricted to axisymmetric
perturbations for simplicity.

There are at least two observations to make about equation
(\ref{eq:curvature}). Firstly, we can see that there is no
physical divergence at the point $r=3\mu/m$, at which the
Schr\"odinger potential diverges. Given that $\Phi$ must go to
zero at this point, the integral in (\ref{eq:curvature}) is
finite. Secondly, taking the limit $r\to 0$ we find
\be
\delta \left( C_{abcd} C^{abcd} \right) \sim a_0 \left(
\frac{1}{r^6}\log\frac{r}{\mu}  + \cdots
\right) + b_0 \left( \frac{1}{r^6} + \cdots
\right) \quad \text{as} \quad r\to 0\,.
\ee
Therefore we see that if $a_0=0$ the perturbed curvature has the
same degree of divergence as the background, whilst if $a_0 \neq
0$ then the divergence of the perturbed curvature is greater than
that of the background.

\subsubsection{Weyl scalars}

Perturbations about black hole spacetime are often considered
using the formalism of Weyl scalars. This is particularly useful
for the case of rotating black holes where the Newman-Penrose
formalism allows a separation of variables of the perturbation
equations \cite{Teukolsky:1972my,Teukolsky:1973ha}. In this
subsection we briefly recast our analysis in terms of a Weyl
scalar.

An explicit transformation is known \cite{chandra,cbook} between
the variable $\Phi$ satisfying the scalar Schr\"odinger equation
(the Zerilli equation) and the perturbed Weyl scalar
\be
\delta\Psi_0 = - \delta C_{pqrs}\, l^{p} m^{q} l^{r}
m^{s} \,,
\ee
where the null vectors are given by
\bea
l & = & \frac{1}{f(r)} \frac{\pa}{\pa t} + \frac{\pa}{\pa r} \,, \nonumber \\
n & = & \frac{1}{2}\frac{\pa}{\pa t} - \frac{f(r)}{2} \frac{\pa}{\pa r} \,, \nonumber \\
m & = & \frac{1}{\sqrt{2}} \left[
\frac{1}{r} \frac{\pa}{\pa \theta} + \frac{i}{r\sin\theta}
\frac{\pa}{\pa \phi} \right]
\,.
\eea

In the Schwarzschild background the only nonvanishing Weyl scalar
is
\be
\Psi_2 = - C_{pqrs}\, l^{p} m^{q} \bar{m}^{r}
n^{s} = \frac{\mu}{2} \frac{1}{r^3}\,.
\ee
The relationship between the perturbed Weyl scalar and the scalar
perturbation, including the angular and time dependence, is
\cite{chandra,cbook}
\bea\label{eq:transform}
\lefteqn{ \delta\Psi_0 = \frac{r}{2(r+\mu)^2} e^{-i\omega
  t}\left(\frac{d^2}{d\theta^2} P_l(\cos\theta) -
\cot\theta \frac{d}{d\theta} P_l(\cos\theta)  \right) \nonumber} \\
& & \times \left[\left(\frac{d}{dr_*}
- i \omega\right) + \frac{2mr^2+3m\mu r-3\mu^2}{r^2(mr-3\mu)} \right]
\left(\frac{d}{dr_*} - i \omega\right) \Phi(r)   \,.
\eea
This equation was derived in \cite{chandra,cbook}
by working in a specific gauge, but given
that $\delta\Psi_0$ is gauge invariant for the Schwarzschild
background, the relation should also hold for our gauge invariant
$\Phi$.

In (\ref{eq:transform}) we can see the appearance of a divergence
at the finite radius $r=3\mu/m$. For the Weyl scalar to remain
bounded at this radius, $\Phi$ must vanish at $r=3\mu/m$. This is
precisely the condition that we found previously due to a
divergence in the scalar potential. Therefore, we see explicitly
that the unphysical divergence in the potential arises in the
change of variables relating the metric perturbation variable to
physical curvatures.

If we consider the $r\to 0$ limit of $\Phi$ (\ref{eq:0sol}) then we
find that the Weyl scalar perturbation behaves as
\be
\delta\Psi_0 \sim a_0\left(\frac{1}{r_*} + \cdots \right) + b_0 \left(
1 + \cdots \right) \quad \text{as} \quad r_* \to 0\, .
\ee

\section{Discussion}

The linearised dynamics of gravitational perturbations about the negative
mass Schwarzschild spacetime are given a well defined dynamics by
specifying boundary conditions at the singularity. There is a one
parameter choice of possible boundary conditions corresponding to
self-adjoint extensions of the Hamiltonian. We considered the
stability of the spacetime as a function of the boundary
condition. We have shown that there is a critical boundary
condition separating stable and unstable spacetimes.

Amongst the possible boundary conditions, we have seen that there
is one particular choice, $a_0=0$ in (\ref{eq:0sol}), which gives
perturbations with finite physical energies $E_1$ and $E_2$.
Furthermore, these perturbations induce curvature perturbations
with the same degree of divergence at the singularity as the
background curvature. All the other boundary conditions result in
perturbations with infinite energy which have an enhanced
curvature divergence at the singularity. Therefore, the $a_0=0$
boundary seems to be more physical than the others. We saw that
this boundary condition gave a stable spacetime.

The main conclusion following from these results is that the
negative mass Schwarzschild phantom can be perturbatively stable
and in particular is stable with what appears to be the most
physical choice of boundary conditions. This is perhaps counter to
intuition, given the existence of nonlinear instabilities that we
reviewed in the introduction.

There are two immediate limitations of our calculations. Firstly, very
near to the singularity it is likely that a linearised analysis of
perturbations is not valid. This is because higher order terms in the
equations of motion will contain curvature tensors that diverge as
$r_* \to 0$. By making the perturbations sufficiently small one might
hope that the linearised approximation will be valid down to very
small radii. Perhaps in this case the nonlinear effects near the
singularity could be absorbed into an effective boundary
condition. Indeed it would be interesting if a preferred boundary
condition is selected in this way.

Secondly, near the singularity the curvatures become large and it
seems likely that quantum corrections to the spacetime will be
important. Again, such corrections could perhaps be renormalised
into an effective boundary condition. In this connection, it is
interesting to note that there are arguments suggesting that the
negative mass Schwarzschild singularity should not be resolved by
quantum corrections \cite{Horowitz:1995ta}.

We would now like to return to the issue of the $l=1$ scalar mode.
As mentioned above, this is a `translational zero mode' which
could be interpreted as allowing the singularity to move from its
initial position. Because the mass of the singularity is negative,
one expects such a motion to carry negative kinetic energy. This
could clearly lead to various nonlinear instabilities including
that envisioned by Bondi, which we reviewed in the introduction.

There are various possible extensions of our calculations.
Adapting to higher dimensional spacetimes should be
straightforward, the necessary formalism may be found in
\cite{Kodama:2003jz,Ishibashi:2003ap,Gibbons:2002pq}. Back in four
dimensions it may be possible to extend the calculations to the
case of rotating Kerr black holes. Remarkably, the linearised
stability of positive mass Kerr black holes is a tractable problem
and stability has been proven
\cite{Teukolsky:1972my,Teukolsky:1973ha,cbook,Whiting:1988vc}.
Negative mass rotating black holes would presumably also have a
negative moment of inertia and therefore potentially suffer from
additional `spin up' instabilities.

More speculatively, there is an intriguing similarity between the
behaviour of the perturbations we have considered near the
singularity and the asymptotic behaviour of perturbations about
Anti-de Sitter space \cite{Ishibashi:2004wx}. Perhaps there is a
version of `holographic renormalisation'
\cite{Balasubramanian:1999re,Kraus:1999di,deHaro:2000xn} which
applies to the naked singularity and allows one to absorb some
divergences arising from integrations near the singularity?

\bigskip
\begin{center}
{\bf Acknowledgements}
\end{center}

We would like to thank Bernard Whiting for some helpful correspondence.
S.A.H. is supported by a Research Fellowship from Clare College,
Cambridge. A.I.'s research is supported in part by the Japan Society
for the Promotion of Science.

\end{document}